\documentclass[preprint]{emulateapj}

\usepackage{graphicx}        
\usepackage{bm}              
\usepackage{natbib}
\usepackage{mathrsfs}

\newcommand{\arone}{148\,GHz}
\newcommand{\artwo}{218\,GHz}

\newcommand{\commentx}[1]{}

\renewcommand{\vec}[1]{\mbox{\boldmath$#1$}} 

\newcommand{\etal}{et al.\,}  

\newcommand{\ra}[3]   
   {\makebox[1.5em][r]{#1}\makebox[1.5em][r]{#2} \makebox[2em][r]{#3}}

\newcommand{\hms}[3]  
   {${#1}^{\mathrm{h}}{#2}^{\mathrm{m}}{#3}^{\mathrm{s}}$}

\newcommand{\hmin}[2]  
   {\ensuremath{{#1}^{\mathrm{h}}{#2}^{\mathrm{m}}}}
   
\newcommand{\hours}[1]  
   {\ensuremath{{#1}^{\mathrm{h}}}}

\newcommand{\dms}[3]  
   {\ensuremath{{#1}\degree{#2}\arcminute{#3}\arcsecond}}

\newcommand{\dm}[2]  
   {\ensuremath{{#1}\degree{#2}\arcminute}}

\newcommand{\ukcmb}  
           {\ensuremath{\micro \kelvin_\mathrm{cmb}}}

\newcommand{\uk}  
           {\ensuremath{\micro \kelvin}}

\newcommand{\fdeg} 
           {\hbox{$.\!\!^{\circ}$}}

\usepackage{color}

\usepackage[mediumspace,Gray,squaren]{SIunits}
\usepackage{rotating}

\newcommand{\npatch}{12~}

\newcommand{\ndivtext}{four }
\newcommand{\ndivword}{a quarter }

\newcommand{\be}{\begin{equation}}
\newcommand{\ee}{\end{equation}}
\newcommand{\bea}{\begin{eqnarray}}
\newcommand{\eea}{\end{eqnarray}}

\shorttitle{ACT Calibration}
\shortauthors{A. Hajian \etal}

\begin{document}

\title{The Atacama Cosmology Telescope: Calibration with WMAP Using Cross-Correlations}
\author{
Amir~Hajian\altaffilmark{1,2,3},
Viviana~Acquaviva\altaffilmark{4,2},
Peter~A.~R.~Ade\altaffilmark{5},
Paula~Aguirre\altaffilmark{6},
Mandana~Amiri\altaffilmark{7},
John~William~Appel\altaffilmark{3},
L.~Felipe~Barrientos\altaffilmark{6},
Elia~S.~Battistelli\altaffilmark{11,7},
J.~Richard~Bond\altaffilmark{1},
Ben~Brown\altaffilmark{12},
Bryce~Burger\altaffilmark{7},
Jay~Chervenak\altaffilmark{13},
Sudeep~Das\altaffilmark{14,3,2},
Mark~J.~Devlin\altaffilmark{9},
Simon~R.~Dicker\altaffilmark{9},
W.~Bertrand~Doriese\altaffilmark{15},
Joanna~Dunkley\altaffilmark{16,3,2},
Rolando~D\"{u}nner\altaffilmark{3},
Thomas~Essinger-Hileman\altaffilmark{3},
Ryan~P.~Fisher\altaffilmark{3},
Joseph~W.~Fowler\altaffilmark{3},
Mark~Halpern\altaffilmark{7},
Matthew~Hasselfield\altaffilmark{7},
Carlos~Hern\'andez-Monteagudo\altaffilmark{17},
Gene~C.~Hilton\altaffilmark{15},
Matt~Hilton\altaffilmark{18,19},
Adam~D.~Hincks\altaffilmark{3},
Ren\'ee~Hlozek\altaffilmark{16},
Kevin~M.~Huffenberger\altaffilmark{20},
David~H.~Hughes\altaffilmark{21},
John~P.~Hughes\altaffilmark{4},
Leopoldo~Infante\altaffilmark{6},
Kent~D.~Irwin\altaffilmark{15},
Jean~Baptiste~Juin\altaffilmark{6},
Madhuri~Kaul\altaffilmark{9},
Jeff~Klein\altaffilmark{9},
Arthur~Kosowsky\altaffilmark{12},
Judy~M~Lau\altaffilmark{22,23,3},
Michele~Limon\altaffilmark{24,9,3},
Yen-Ting~Lin\altaffilmark{25,2,6},
Robert~H.~Lupton\altaffilmark{2},
Tobias~A.~Marriage\altaffilmark{2,8},
Danica~Marsden\altaffilmark{9},
Phil~Mauskopf\altaffilmark{5},
Felipe~Menanteau\altaffilmark{4},
Kavilan~Moodley\altaffilmark{18,19},
Harvey~Moseley\altaffilmark{13},
Calvin~B~Netterfield\altaffilmark{27},
Michael~D.~Niemack\altaffilmark{15,3},
Michael~R.~Nolta\altaffilmark{1},
Lyman~A.~Page\altaffilmark{3},
Lucas~Parker\altaffilmark{3},
Bruce~Partridge\altaffilmark{10},
Beth~Reid\altaffilmark{28,3},
Neelima~Sehgal\altaffilmark{22},
Blake~D.~Sherwin\altaffilmark{3},
Jon~Sievers\altaffilmark{1},
David~N.~Spergel\altaffilmark{2},
Suzanne~T.~Staggs\altaffilmark{3},
Daniel~S.~Swetz\altaffilmark{9,15},
Eric~R.~Switzer\altaffilmark{26,3},
Robert~Thornton\altaffilmark{9,29},
Hy~Trac\altaffilmark{30,31},
Carole~Tucker\altaffilmark{5},
Ryan~Warne\altaffilmark{18},
Ed~Wollack\altaffilmark{13},
Yue~Zhao\altaffilmark{3}
}
\altaffiltext{1}{Canadian Institute for Theoretical Astrophysics, University of Toronto, Toronto, ON, Canada M5S 3H8}
\altaffiltext{2}{Department of Astrophysical Sciences, Peyton Hall, 
Princeton University, Princeton, NJ USA 08544}
\altaffiltext{3}{Joseph Henry Laboratories of Physics, Jadwin Hall,
Princeton University, Princeton, NJ, USA 08544}
\altaffiltext{4}{Department of Physics and Astronomy, Rutgers, 
The State University of New Jersey, Piscataway, NJ USA 08854-8019}
\altaffiltext{5}{School of Physics and Astronomy, Cardiff University, The Parade, Cardiff, Wales, UK CF24 3AA}
\altaffiltext{6}{Departamento de Astronom{\'{i}}a y Astrof{\'{i}}sica, 
Facultad de F{\'{i}}sica, Pontific\'{i}a Universidad Cat\'{o}lica de Chile,
Casilla 306, Santiago 22, Chile}
\altaffiltext{7}{Department of Physics and Astronomy, University of
British Columbia, Vancouver, BC, Canada V6T 1Z4}
\altaffiltext{8}{Dept. of Physics and Astronomy, The Johns Hopkins University, 3400 N. Charles St., Baltimore, MD 21218-2686}
\altaffiltext{9}{Department of Physics and Astronomy, University of
Pennsylvania, 209 South 33rd Street, Philadelphia, PA, USA 19104}
\altaffiltext{10}{Department of Physics and Astronomy, Haverford College,
Haverford, PA, USA 19041}
\altaffiltext{11}{Department of Physics, University of Rome ``La Sapienza'', 
Piazzale Aldo Moro 5, I-00185 Rome, Italy}
\altaffiltext{12}{Department of Physics and Astronomy, University of Pittsburgh, Pittsburgh, PA, USA 15260}
\altaffiltext{13}{Code 553/665, NASA/Goddard Space Flight Center,
Greenbelt, MD, USA 20771}
\altaffiltext{14}{Berkeley Center for Cosmological Physics, LBL and
Department of Physics, University of California, Berkeley, CA, USA 94720}
\altaffiltext{15}{NIST Quantum Devices Group, 325
Broadway Mailcode 817.03, Boulder, CO, USA 80305}
\altaffiltext{16}{Department of Astrophysics, Oxford University, Oxford, 
UK OX1 3RH}
\altaffiltext{17}{Max Planck Institut f\"ur Astrophysik, Postfach 1317, 
D-85741 Garching bei M\"unchen, Germany}
\altaffiltext{18}{Astrophysics and Cosmology Research Unit, School of
Mathematical Sciences, University of KwaZulu-Natal, Durban, 4041,
South Africa}
\altaffiltext{19}{Centre for High Performance Computing, CSIR Campus, 15 Lower
Hope St. Rosebank, Cape Town, South Africa}
\altaffiltext{20}{Department of Physics, University of Miami, Coral Gables, 
FL, USA 33124}
\altaffiltext{21}{Instituto Nacional de Astrof\'isica, \'Optica y 
Electr\'onica (INAOE), Tonantzintla, Puebla, Mexico}
\altaffiltext{22}{Kavli Institute for Particle Astrophysics and Cosmology, Stanford
University, Stanford, CA, USA 94305-4085}
\altaffiltext{23}{Department of Physics, Stanford University, Stanford, CA, 
USA 94305-4085}
\altaffiltext{24}{Columbia Astrophysics Laboratory, 550 W. 120th St. Mail Code 5247,
New York, NY USA 10027}
\altaffiltext{25}{Institute for the Physics and Mathematics of the Universe, 
The University of Tokyo, Kashiwa, Chiba 277-8568, Japan}
\altaffiltext{26}{Kavli Institute for Cosmological Physics, 
Laboratory for Astrophysics and Space Research, 5620 South Ellis Ave.,
Chicago, IL, USA 60637}
\altaffiltext{27}{Department of Physics, University of Toronto, 
60 St. George Street, Toronto, ON, Canada M5S 1A7}
\altaffiltext{28}{ICREA \& Institut de Ciencies del Cosmos (ICC), University of
Barcelona, Barcelona 08028, Spain}
\altaffiltext{29}{Department of Physics , West Chester University 
of Pennsylvania, West Chester, PA, USA 19383}
\altaffiltext{30}{Harvard-Smithsonian Center for Astrophysics, 
Harvard University, Cambridge, MA, USA 02138}
\altaffiltext{31}{Department of Physics, Carnegie Mellon University, Pittsburgh, PA 15213}

\begin{abstract}
We present a new calibration method based on cross-correlations with WMAP and apply it to data from the Atacama Cosmology Telescope (ACT). ACT's observing strategy and map making procedure allows an unbiased reconstruction of the modes in the maps over a wide range of multipoles. By directly matching the ACT maps to WMAP observations in the multipole range of $400 < l < 1000$, we determine the absolute calibration with an uncertainty of $2\%$ in temperature. The precise measurement of the calibration error directly impacts the uncertainties in the cosmological parameters estimated from the ACT power spectra.   We also present a combined map based on ACT and WMAP data that has high signal-to-noise over a wide range of multipoles. 
\end{abstract}
 

\section{INTRODUCTION}

\setcounter{footnote}{0}

Observations of the cosmic microwave background (CMB) are one of the key probes of cosmology.  Current balloon and ground based measurements \citep{sayers/etal:2009,
brown/etal:2009, friedman/etal:2009,
reichardt/etal:2009a,
reichardt/etal:2009,
sharp/etal:2010,
dawson/etal:2006,
sievers/etal:prep,
lueker/etal:2010,
fowler/etal:prep}   
   and space-based experiments \citep{komatsu/etal:prep}  measure the amplitude of cosmic microwave background fluctuations over a wide range of angular scales. All-sky maps of the Wilkinson Microwave Anisotropy Probe (WMAP) provide a large-scale measurement of the cosmic microwave background with only a 0.2\% calibration uncertainty \citep{jarosik/etal:prep}.   The scientific value of the small-scale ground and balloon-based measurements are maximized if these precise small-scale maps can be accurately calibrated to WMAP. 

Starting with the comparisons between FIRS and COBE \citep{ganga/etal:1993}, 
cross-calibration studies have both confirmed detections and provided common calibration between experiments.  General methods for cross-calibration have been developed  for CMB studies \citep{ganga/etal:1993, tegmark:1999} and applied to many different experiments: \cite{reichardt/etal:2009}, (ACBAR with WMAP), \cite{brown/etal:2009}, (QUaD with BOOMERang), \cite{chiang/etal:prep}, (BICEP with WMAP), \cite{xu/etal:2001},  (SASK and QMAP with COBE),  \cite{netterfield/etal:1996}, (SASK with MSAM), \cite{abroe/etal:2003}, (Maxima with WMAP), \cite{hernandez-monteagudo/etal:2005}, (ARCHEOPS and WMAP), \cite{masi/etal:2006}, (BOOMERanG and WMAP). 
  An alternative approach to calibration relies on measuring known point sources, particularly planets. This approach has also been used by many CMB experiments  \citep{crill/etal:2003, mason/etal:2003} and is applied in section \ref{subsec:planets}.

We present a new method based on cross-correlations with WMAP to accurately measure the absolute calibration uncertainty for ACT observations.  The method we present here expands on earlier work in that we take full advantage of working in Fourier space. The Fourier space algorithms introduced in this paper are faster than the real-space algorithms. They are simple and scale linearly with the number of pixels in the maps. Filtering noisy modes, deconvolving the beam and estimating the noise model is faster and easier in Fourier space. The ACT data used in this paper were collected at \arone{} and \artwo{} during  the 2008 observing season. The maps are solved for iteratively using a preconditioned conjugate gradient code to obtain the unbiased maximum likelihood solution. For an overview of the ACT data reduction pipeline, we refer the reader to \cite{fowler/etal:prep, dunner:2009}. For the power spectrum results and cosmological parameters, see the companion papers \citep{das/etal:prep, dunkley/etal:prep}, and for clusters using the Sunyaev-Zel'dovich effect \citep{sunyaev/zeldovich:1970} see \cite{marriage/etal:prepb} and \cite{menanteau/etal:prep}.  For information on the telescope facility, see \cite{fowler/etal:2007, hincks/etal:2008, switzer/etal:2008, swetz/etal:prep}. Details of the cryogenic receiver and bolometric detectors are provided in \cite{niemack:2006, marriage/chervenak/doriese:2006, fowler/etal:2007, battistelli/etal:2008, niemack/etal:2008, swetz/etal:2008, thornton/etal:2008, zhao/etal:2008}.

In Section \ref{sec:data} we briefly describe the data sets used for this analysis. In Section \ref{sec:alignment} we present a cross-correlation method to test the relative alignment of the ACT maps versus WMAP maps. Our calibration method is discussed in Section \ref{sec:calibration}. Section \ref{sec:optimalMap} presents our combined high resolution CMB map that is signal-to-noise dominated over a wide range of multipoles ($l<5000$).

\begin{figure}[t]
\centering
     \plotone{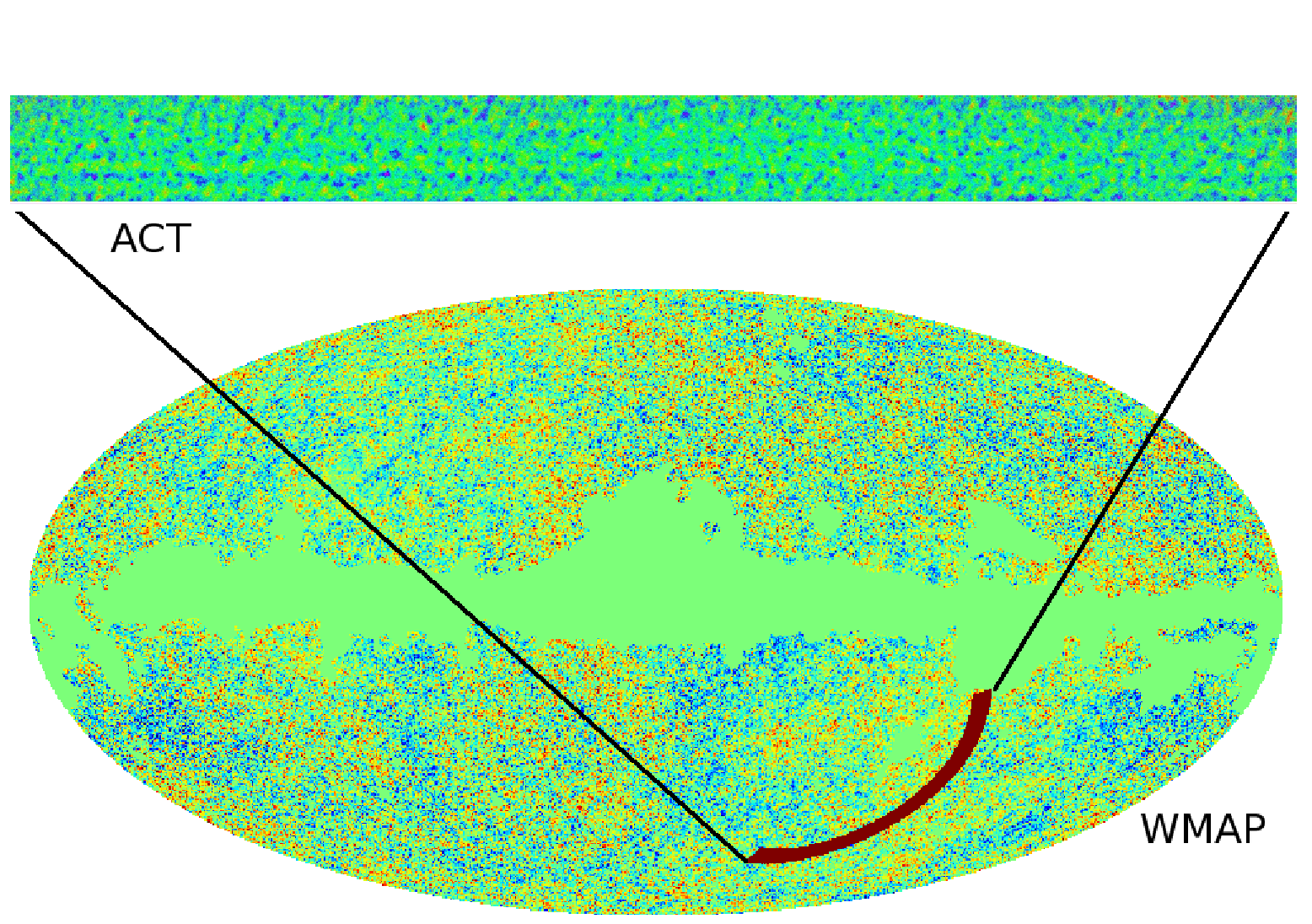}
  \caption{The ACT and WMAP maps used in this paper. The red region in the WMAP map shows the overlap between WMAP and the ACT data.}
  \label{fig:ACTonWMAP}
\end{figure}        

\section{Data}
\label{sec:data}
The cross-correlation method used in this paper requires a single map from WMAP and four maps with independent noise properties from ACT at each frequency. Below we give a brief description of the three data sets that are used in this paper. Throughout this paper we use  $a$  to denote ACT data and $w$ to denote WMAP data. 
 
\subsection{ACT Data}
One of the strengths of ACT is its scanning strategy. The geographical location of the ACT telescope (the Atacama desert) enables cross-linking of observations. Every observing region is scanned along two different directions at constant elevation each night. Azimuth-only scans observe a sky field  at the same elevation twice a night: once as the field rises; once as it sets. Sky rotation changes the scan angle, resulting in cross-linking, which allows (in principle) the reconstruction of all the modes in the map. The modes that are lost due to the correlated noise along the scan direction in the rising maps are won back using the setting maps and vice versa. Such cross-linking and the unbiased map-making algorithm used to make ACT maps yield a successful reconstruction of the modes in the maps down to multipoles of $\sim 300$. The combination of ACT's scan strategy and its unbiased map-making scheme enables a direct calibration to WMAP.  However, since WMAP is noisy on small-scales and ACT's maps are dominated by atmospheric noise and poorly measured modes on its largest scales, our method compares the maps as a function of multipole over the $400 < l < 1000$ range.

The ACT maps used in this paper are identical to the maps used in \cite{das/etal:prep}. The area includes that in \cite{fowler/etal:prep} and represents the highest sensitivity data from 2008 (\arone{} and \artwo{}). 
 The map resolution is $1.4\arcmin$ and $1.0\arcmin$ at \arone{} and \artwo{}, respectively \citep{hincks/etal:prep}. The map projection used is cylindrical equal area with square pixels, $0.5\arcmin$ on a side. 
 We divide the data into \npatch patches. Each patch is $5 \degree \times 5 \degree$ in size, and together they cover a rectangular area of the map from $\alpha =$ $\hmin{00}{22}$  to  $\hmin{06}{52}$
(5\fdeg 5 to  103\degree) in right ascension and from $\delta = {-55}\degree$  to ${-50}\degree$ in declination.
 We divide our data-set into \ndivtext equal
subsets in time, such that the \ndivtext  independent maps generated from these subsets cover the same area and have similar depths. Therefore, for each
patch described above we have \ndivtext maps, each representing roughly
\ndivword of the time spent on that patch. We call these ``season maps''.

\subsection{Calibration to Planets}
\label{subsec:planets}
Observations of Uranus provide the initial calibration of the maps. Uranus was observed by ACT every few days during the 2008 season, yielding approximately 30 usable observations.  The time-ordered data from each observation are calibrated to detector power units and a map is produced.  From each map, the peak response of the planet is determined, corrected for the temperature dilution due to the finite instrumental beam size (\textit{i.e.} by multiplying by the ratio of Uranus's solid angle to the instrumental beam solid angle)  
 and then compared to the Uranus temperature (converted to CMB differential units) at the band effective frequency.  The calibration factors from detector power units to CMB differential temperature are compared to precipitable water vapor (PWV) measurements to fit a model of atmospheric opacity, and the season map calibration factor is obtained by evaluating the fit at the season mean PWV of 0.5.  The uncertainty in the Uranus-based calibration is $7\%$, and is dominated by the $5\%$ uncertainty in the planet's temperature.

We take the brightness temperature\footnote{Throughout, the \emph{brightness temperature} refers to the temperature of a black-body having the observed in-band spectral radiance.  The equivalent \emph{Rayleigh-Jeans temperatures} of Uranus for our bands are 103.5 and 90.7 K.} of Uranus to be $107 \pm 6$ K and $96 \pm 6$ K for the \arone{} and \artwo{} bands respectively.  These temperatures are based on a reprocessing of the data presented by \cite{griffin/orton:1993}, in combination with WMAP 7-year measurements of Mars and Uranus brightnesses \citep{weiland/etal:inprep}.  Griffin and Orton measured flux ratios between Uranus and Mars at several wavelengths from 350 $\micro$m to $1.9$ mm.  These were calibrated to absolute units using an extrapolation of the Wright model \citep{wright:1976} for Mars temperature, and the resulting Uranus brightness temperatures were fit with a third-order polynomial in the logarithm of the wavelength.  We have followed the same procedure, but with two modifications.  The first is that the WMAP measurements of Uranus temperature at 94, 61 and 41 GHz have been included in the fit. The second is that the WMAP comparison of Mars temperature at 94 GHz (3.2 mm) to the Wright model at 350 $\micro$m is used to pin the long wavelength end of the Mars brightness extrapolation.  The resulting Uranus brightness temperatures in the two bands are $5\%$ and $3\%$ lower than those obtained by Griffin and Orton.  This difference is due almost entirely to the recalibration of the Mars model.

\begin{figure*}[htb]
\centering
    \resizebox{.7\textwidth}{!}{
     \plotone{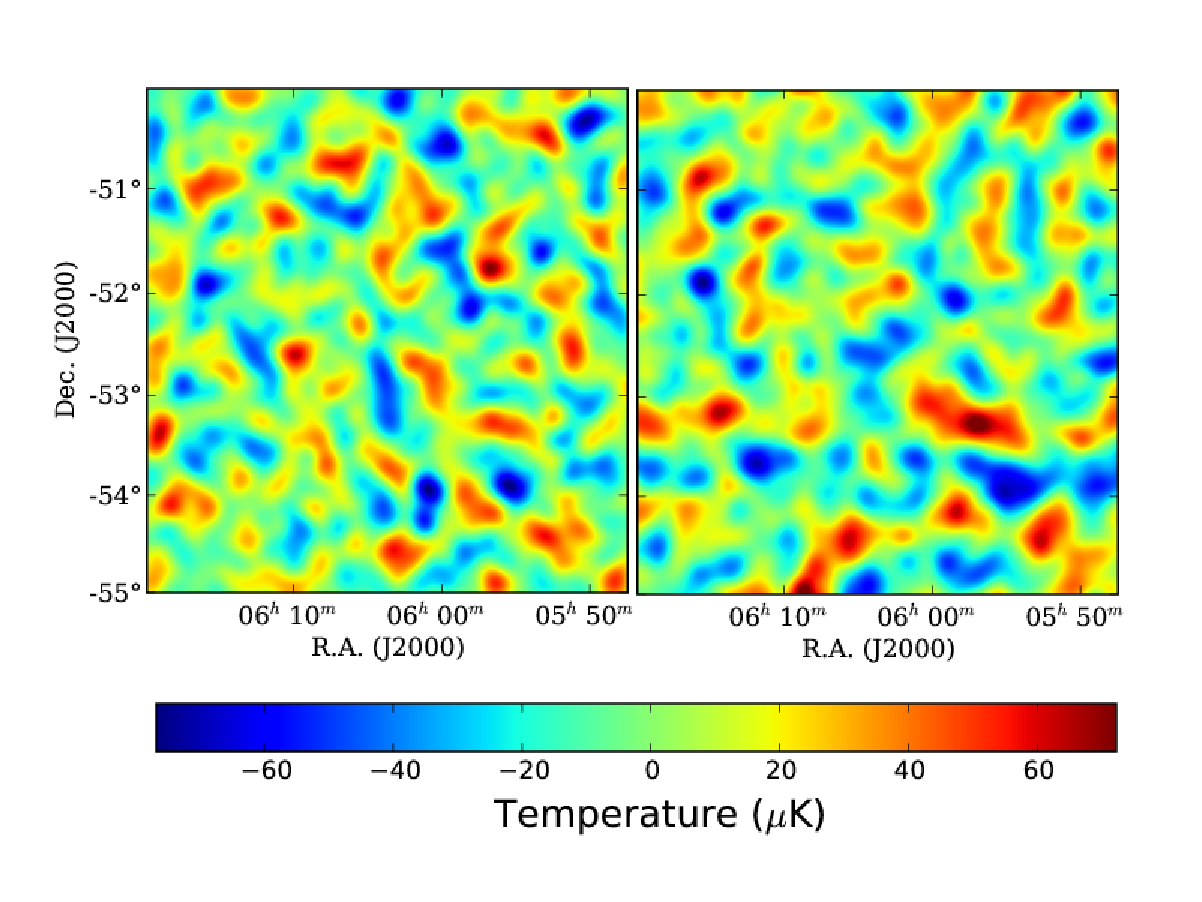}
}
  \caption{ One of the \npatch patches used for cross-correlations from ACT \arone{} (\textit{right}),  and WMAP data (\textit{left}). Long wavelength modes have been removed from both maps using the high-pass filter described in eqn. (\ref{eqn:cos2filter}). The ACT map has been convolved with the WMAP W-band beam. The two maps represent the same area in the sky, observed by two experiments. The similarity between structures in the two maps is the key feature that we use for our cross-correlation studies. The ACT maps have large-scale atmospheric noise and the WMAP maps are dominated by detector noise on the smallest scales.  These two types of noise are completely uncorrelated.}
  \label{fig:wMapVsACT}
\end{figure*}

\subsection{WMAP Data}
The WMAP map uses co-added inverse-noise weighted data from seven single year maps and four differencing assemblies at $94$ GHz.\footnote{from \url{http://lambda.gsfc.nasa.gov}} 
The maps are foreground cleaned (using the foreground template model discussed in \cite{hinshaw/etal:2007}) and are at HEALPix \footnote{\url{http://healpix.jpl.nasa.gov}}  resolution 10 ($N_{\rm side}=1024$), with $3.5\arcmin$ pixels. Single year maps are multiplied by the pixel weights based on pixel noise evaluated with the expression
\be
\sigma^2(\vec n) = \sigma^2_W/N_{obs},
\ee
where $N_{obs}$ is the number of observations at each map pixel which is directly proportional to the statistical weight and $\sigma_W$ is the noise for each differencing assembly. The four noise factors in mK units are given in Table \ref{table}.

\begin{table}
\large
\caption{The rms noise for each WMAP differencing assembly.}
\label{table}
\normalsize
\begin{tabular*}{\columnwidth}{@{\extracolsep{\fill}}c|cccc}
\hline
\hline
 &$W_1$ & $W_2$ & $W_3$ & $W_4$\\ 
\hline
$\sigma_W$(mK) &5.906 & 6.572& 6.941 & 6.778\\
\hline
\hline
\end{tabular*}
\label{table_1}
\end{table}

We cut a rectangular region from the resulting HEALPix map corresponding to the ACT southern strip and project it onto the CEA coordinates.
Both the WMAP and ACT data are cut into the same \npatch patches. Figure \ref{fig:wMapVsACT} shows one of these patches and the corresponding patch from the ACT measurements. While both maps shows similar hot and cold spots, the WMAP beam has smoothed out small scale structure and the ACT maps contain large-scale atmospheric noise.

Because the ACT maps have poorly measured modes on the largest scales, the WMAP and ACT maps are filtered by a high-pass filter $F_c(l)$ in Fourier space before being cut into \npatch patches. The high-pass filter is a smooth sine-squared function in Fourier space given by
\be
\label{eqn:cos2filter}
F_c(l) = \sin^2{x(l)} \Theta(l-l_{\mathrm{min}}) \Theta(l_{\mathrm{max}}-l) + \Theta(l-l_{\mathrm{max}}),
\ee
where $x(l)=(\pi/2)(l-l_{\mathrm{min}})/(l_{\mathrm{max}}-l_{\mathrm{min}})$ and $\Theta$ is the Heavyside function. We choose $l_{\mathrm{min}} = 100$ and $l_{\mathrm{max}}=500$. The final power spectrum is corrected for this filter as well as for the effects of the beam and pixel window functions.

\section{Comparing Maps}
\label{sec:alignment}
\begin{figure*}[htb]
\centering
    \resizebox{\textwidth}{!}{
    \plotone{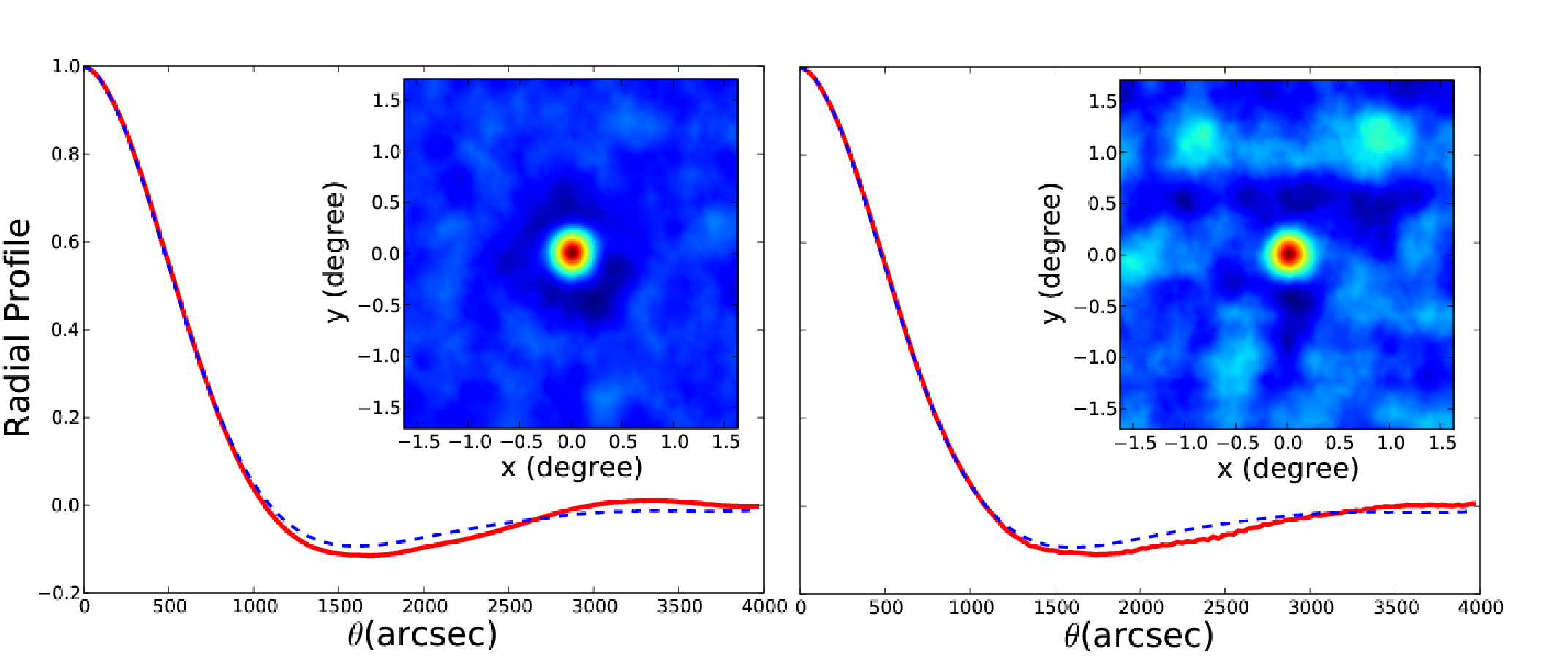}
}
  \caption{Comparing ACT and WMAP data sets: Two-dimensional cross-correlations of WMAP and ACT maps in real-space (\textit{inset panels}) and their one dimensional radial profiles (\textit{solid red curves}) for the \arone{} maps  (\textit{left}) and \artwo{} maps (\textit{right}).  2D cross-correlations peak at zero lag. This solidifies that the maps are aligned and the pointing agrees on average over the whole map area. Results obtained from the noiseless simulations of \cite{sehgal/etal:2010} are also plotted (\textit{dashed blue lines}). The data correlation functions agree with the simulations. A  $\sigma=7.7'$  Gaussian profile is a good fit to the 2D correlations at small scales ($<10'$).   }
  \label{fig:alignment}
\end{figure*}
We test the alignment of the maps by computing their real-space cross-correlation function as a function of offset $\vec{ x}$.   This is done by inverse Fourier transforming the 2-dimensional cross-correlation in Fourier space:
\be
 M_{a\times w}(\vec x) = \sum_{\vec l} \tilde a(\vec l) \tilde w^*(\vec l) \exp(i \vec l \cdot \vec x).
\ee
Here $ a(\vec l) $ and $ w(\vec l) $ are the Fourier transforms of the ACT and WMAP maps respectively, and the tilde denotes high-pass filtering of those maps, $\tilde a(\vec l) = a(\vec l)F_c(\vec l)$ and $\tilde w(\vec l) = w(\vec l)F_c(\vec l)$. The high-pass filtering is important here because otherwise long wavelength noisy modes dominate the cross correlation and hide the true signal.
The resulting $2$D cross-correlation function, $M_{a\times w}$, can be written as
\be\label{eqn:M}
 M_{a\times w}(\vec x) = C(\vec x)\otimes B(\vec x) + n(\vec x),
\ee
where $B(\vec x)$ is an effective beam between the two maps, $C(\vec x)$ is the correlation function of the two sky maps and $n(\vec x)$ is the noise fluctuations in the cross-correlation that gets smaller as we add more area.  For a perfectly aligned pair of maps, the $2$D cross-correlation function, $M_{a\times w}(\vec x)$ peaks at $\vec x = 0$ and its width is determined by the correlation length of the field and the beams. For our maps, the result of this is shown in Figure \ref{fig:alignment}. The cross-correlation function is cleanly peaked at zero lag, which shows the maps are aligned and correlated. The noiseless sky simulations of \cite{sehgal/etal:2010} provide a confirmation of these results.  
We convolve the simulated maps with the WMAP and ACT beams to generate simulated noiseless maps of WMAP and ACT data respectively. The maps are then filtered with the high-pass filter of eqn. (\ref{eqn:cos2filter}) and cut into \npatch patches. The real-space cross-correlation of the simulated maps is computed using eqn. (\ref{eqn:M}). The result is plotted in Figure \ref{fig:alignment} (dashed blue lines). The agreement between cross-correlations computed from the data and the simulations is striking. We fit a Gaussian curve to the radial profiles of the 2D cross-correlations. At small scales ($<10\arcmin$) a  $\sigma=7.7\arcmin$  Gaussian profile is a good fit, and the result of the simulations agrees well with the data.

\section{Calibration using cross correlations}
\label{sec:calibration}
This paper uses four kinds of cross-power spectra, ACT$\times$WMAP, ACT$\times$ACT ACT$_{148}\times$ACT$_{218}$, and WMAP$\times$WMAP to provide CMB-based calibrations of the ACT maps. The advantage of working with the cross-power spectrum is that the noise in the two maps is independent and therefore uncorrelated.
 Hence a precise modeling of the noise is not needed.  

 If the noise in the two maps is uncorrelated, the cross-power spectrum provides an unbiased estimator of the underlying power spectrum. Cross-correlations can be done between pairs of maps from two different experiments. Because the two experiments have different systematics that do not correlate, the cross-spectrum provides a good estimate of the true underlying power spectrum that is nearly independent of the noise properties in the maps. 

\subsection{Power Spectrum Method}
The power spectrum method we use is the Adaptive Multi-Taper Method (AMTM) of \cite{das/hajian/spergel:2009}.  Also we would like to have the maximum resolution possible in Fourier space to have as many bins in the power spectrum as possible, in order to maximize the number of independent measurements of the calibration factor as it is described below. The size of the bins is limited by the fundamental frequency determined by the smallest side of the maps. For the maps we are using, this is $\delta l = 72$.  Using the AMTM with multiple tapers would result in smaller errors at the large $l$ regime of the power spectrum, but it increases the size of the independent bins. For this reason we use AMTM with one taper at resolution $Nres=1$ and thus no iterations are necessary (see \cite{das/hajian/spergel:2009} for the details of the AMTM method). We use slightly larger bin sizes than the fundamental frequency ($\delta l = 90$) to guarantee that the bins are uncorrelated. This is further tested and verified by Monte Carlo simulations. The prewhitening method of \cite{das/hajian/spergel:2009} is designed to reduce the dynamic range of the Fourier components of the maps. This is important when one is interested in measuring the damping tail of the CMB ($l>1000$) where the slope of the power spectrum is steep. However we do not use prewhitening as we are working in the mildly colored regime of the power spectrum, $\l < 1000$ where the dynamic range of the power spectrum is not large. Using prewhitening does not affect the power spectrum at the $l$ ranges of our interest.

The ACT power spectrum (both for \arone{} and \artwo{} maps) is computed using cross correlations of the four season maps. The power spectrum of each patch is the average of the six cross-power spectra:
\be
\label{patch Spectra}
C_l^i = \frac{1}{6}\sum_{\alpha_i, \beta_i; \alpha_i < \beta_i}^{1\le\beta_i\le4}{C_l^{\alpha_i\beta_i}}
\ee  
where $C_l^{\alpha_i\beta_i}$ are cross-power spectra of season maps for the patch $i$, and $\alpha_i$ and $\beta_i$ index the four season maps of that patch.
The final ACT power spectrum is given by the average of the \npatch patches
\be
C_l = \frac{1}{\npatch}\sum_{i=1}^{\npatch} C_l^i.
\ee 
Each of these \npatch power spectra is an independent measurement of the ACT power spectrum. We use the variance from the \npatch power spectra values at each $l$ bin as a measure of the error on the power spectrum. This method agrees well with the analytical estimate of the errors \citep{fowler/etal:prep}.

The ACT$_{148}\times$ACT$_{218}$ cross-power spectrum is computed in a similar way, but in this case  $\alpha_i$ and $\beta_i$ correspond to the season maps from \arone{} and \artwo{} data respectively, and there are 12 cross-spectra for each patch:
\be
C_l^i = \frac{1}{12}\sum_{\alpha_i, \beta_i; \alpha_i < \beta_i}^{1\le\beta_i\le4}{\left(C_l^{\alpha_i\beta_i}+C_l^{\beta_i\alpha_i}\right)}. 
\ee

The ACT$\times$WMAP cross-power spectrum, $C_l^{aw}$, is given by the average of the \npatch-patch cross-power spectra. The spectrum in each patch is given by the average of the four cross-spectra between the WMAP map and each of the four ACT season maps for that patch,
\be
C_l^i = \frac{1}{4}\sum_{\alpha_i=1}^{4}{C_l^{\alpha_i}}
\ee 
where $\alpha_i$ indexes the four season maps of the $i$th patch in ACT data and
\be
C_l^{\alpha_i} = \frac{1}{N_l}\int{\mathrm{d}\theta \tilde{a}_{\alpha_i}(\vec l)  \tilde{w}^*(\vec l)}, 
\ee
where $N_l$ is the number of modes in each bin in Fourier space.

The window function deconvolution of the power spectra is done on the average cross-power spectrum of each patch.
We have  tested our pipeline on 1000 Monte Carlo simulations to confirm that the power spectrum method we use (including the $2$D noise weighting and the $l$-space masking) is unbiased and that the covariance matrix of the power spectrum bins is diagonal. The patch sizes used in \cite{das/etal:prep} are three times larger than the patch sizes used in this paper, but the number of patches in \cite{das/etal:prep} is three times fewer than that used here. So the total areas covered by both cutting methods are identical and for the same binning, the final power spectra obtained from the two methods agree with each other to better than 1\% fractional error.  

\begin{figure*}[htb]
\centering
    \resizebox{.7\textwidth}{!}{
    \plottwo{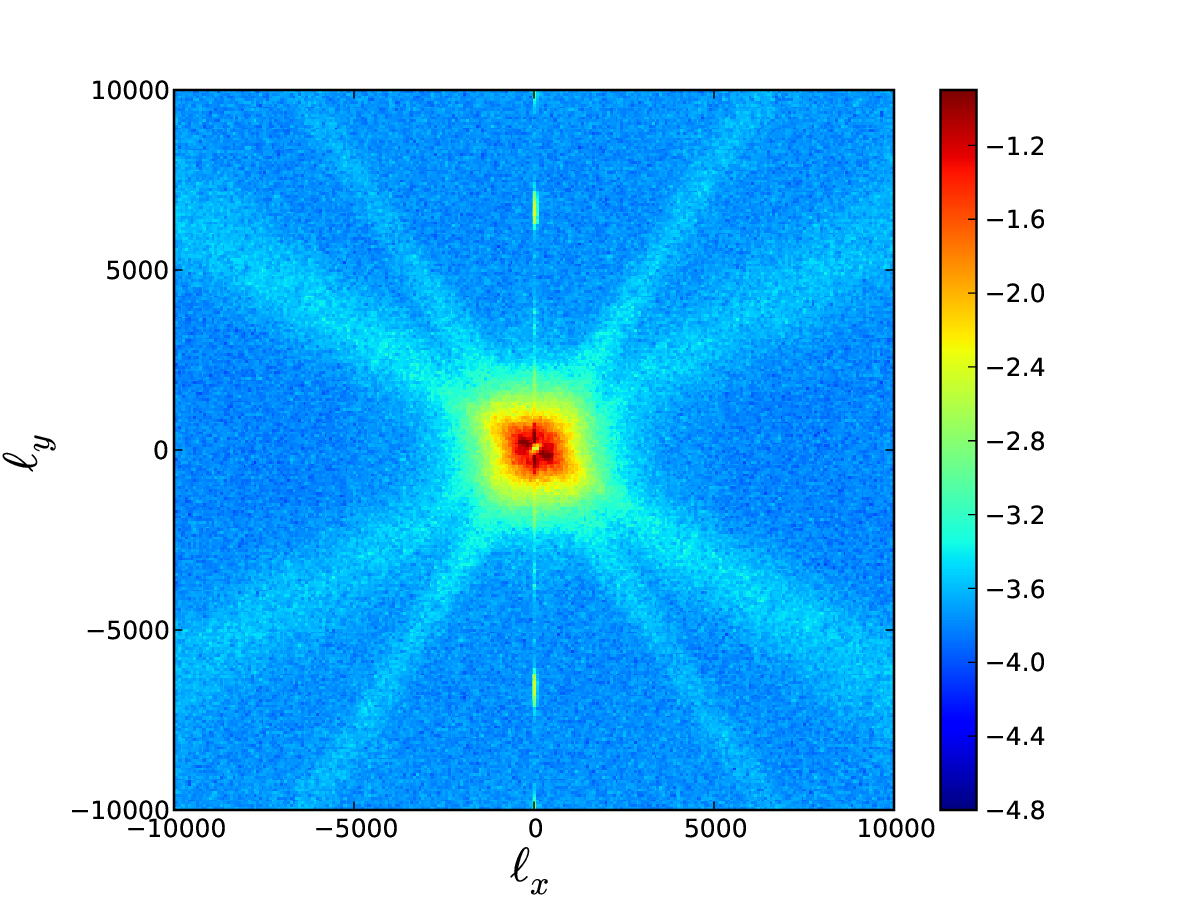}{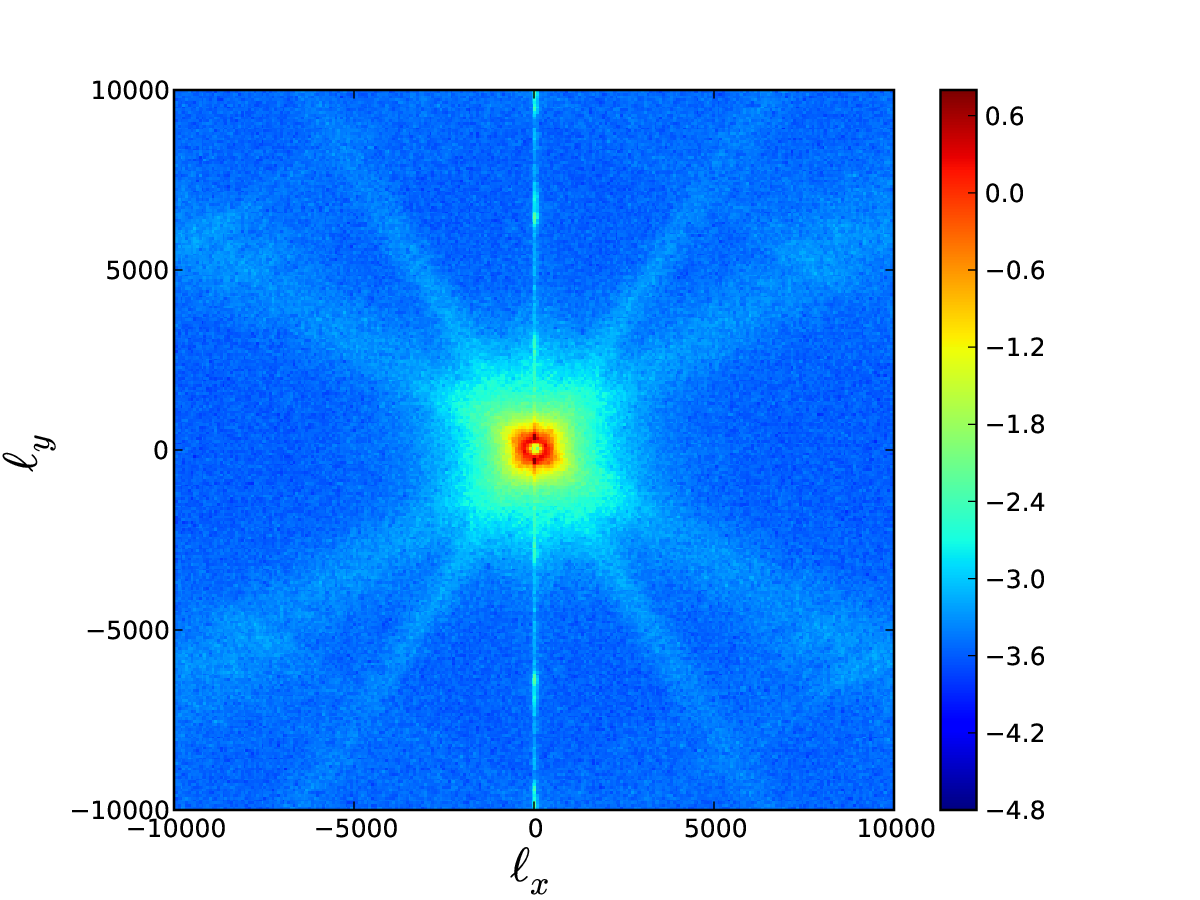}
 
}
  \caption{Average of 72 noise estimates ($\micro K^2$) in 2-dimensions for \arone{} (\textit{left}) and \artwo{} (\textit{right}).    Each estimate is computed from the auto-spectrum of a difference map obtained from differencing two quarter-season maps from the same region of the sky. The sky signal cancels in the difference maps and only noise remains. 
The anisotropy of the noise power spectrum is the reason we use these noise models to down-weight the noisy regions in $2$-dimensional power spectra before binning. These weights are different from those of \cite{marriage/etal:prepb} in that  the CMB is the main signal for us and it does not contribute to the noise model. 
  }
  \label{fig:noise model}
\end{figure*}
The straight binning of the two-dimensional power spectra is the simplest but not the best. Down-weighting noisy regions in the $2$D power spectrum space before binning is a useful technique that is adopted to improve the power spectrum method. Below we discuss our method for estimating the noise model and the $2$D noise weighting along with the $l$-space masking. 

\subsection{Noise Model}
The season maps for each patch share the same signal but have independent noise properties. Differencing the two season maps removes the sky signal and leaves behind a linear combination of the noise in the two maps. 
The resulting map, which we call the ``difference map,'' is a noise map and its auto-power spectrum can be used for estimating the noise model for that patch (see \cite{marriage/etal:prepb} for a more detailed discussion on noise weighting in Fourier space). Figure \ref{fig:noise model} shows the stacked noise models given by the average of 72  two dimensional power spectra obtained from $6$ difference maps per patch, for \npatch patches.  

\subsection{Noise Weighting}
The ACT and WMAP maps have very different noise properties. The WMAP noise spectrum is dominated by the detector noise and is nearly constant over a wide range of the wave-vector $l$, whereas ACT noise has more detailed structure. The ACT noise is dominated by atmospheric noise  on large angular scales, and by the detector noise  on small scales.
Some directions in the two-dimensional ACT power spectrum are more noisy than others. We use our best estimate of the noise for each map to down weight the noisy parts of the spectra before angle-averaging the two-dimensional spectra. The noise for the ACT-ACT cross-spectrum is $N_{aa}^2(\vec{l})$ and for the ACT-WMAP cross-spectrum it is $N_{aa}(\vec{l})N_{ww}(\vec{l})$, where $N_{aa}(\vec{l})$ is the ACT noise spectrum in two dimensions as shown in Figure \ref{fig:noise model} and $N_{ww}(\vec{l})$ is the WMAP noise. Noise weighted power spectra are then angle-averaged and binned in $l$. In the end, the spectra are divided by the beam and relevant pixel window functions. If we denote the two dimensional spectra by $P(\vec{l})$, the final ACT-ACT cross-spectrum is obtained by
\be
C^{aa}(l) =\frac{\int \mathrm{d}\theta \tilde{P}_{aa}(\vec{l})/{N_{aa}^2(\vec{l})}}{F_c^2(l)b_a^2(l) \int \mathrm{d}\theta/{N_{aa}^2(\vec{l})}},
\ee
where $b_a(l)$ is the ACT beam in Fourier space \citep{hincks/etal:prep}, $F_c(l)$ is  the high-pass filter  in Fourier space and $\int d \theta$ represents angle-averaging and binning in $l$. We do not correct for the ACT pixel window function as it is close to unity in the range of interest, $l<1000$.

The ACT-WMAP cross-spectrum is obtained from
\be
C^{aw}(l) =\frac{\int \mathrm{d}\theta \tilde{P}_{aw}(\vec{l})/{N_{aa}(\vec{l})}}{F_c^2(l)b_a(l)b_w(l)p_w(l)\int \mathrm{d}\theta/{N_{aa}(\vec{l})}},
\ee
where $b_w(l)$ is the W band beam of WMAP \citep{jarosik/etal:prep} and $p_w(l)$
is the pixel window function of the resolution 10 HEALPix maps corresponding to  $N_{side} = 1024$, with $3.5'$ pixels. We use HEALPix in Python (healpy\footnote{\url{http://code.google.com/p/healpy}}) to compute the pixel window function at the integer harmonic indices and then bin it in $l$ to get $p_w(l)$. The WMAP noise term gets canceled in the above noise weighting as it is constant.

We mask a vertical band of width $\Delta l_x = 180$ in the power spectra before angle-averaging them. This makes sure that our estimates of the power spectra are not affected by the striping effects that are present at a narrow band around $l=0$. For this analysis, we use a $l$-space mask at $l_x=[-90, 90]$. For a detailed description of the $l$-space masking see \cite{fowler/etal:prep}.

\subsection{Calibration: ACT \arone{}}
We assume that the calibration factor is constant on various scales and that the ACT, $a(x)$, and WMAP , $w(x)$, maps can be represented as
\bea
a(\vec x) &=& \alpha^{-1} \Delta T_{sky}(\vec x)\otimes B_a(\vec x) + N_a(\vec x),\\ \nonumber
w(\vec x) &=& \Delta T_{sky}(\vec x)\otimes B_w(\vec x) + N_w(\vec x),
\eea
where $\alpha$ is the calibration factor \citep{dunner:2009}, $ \Delta T_{sky}(\vec x)$ is the sky temperature signal, $N_a$  and $N_w$ are ACT and WMAP noises respectively, $B_a$ is the ACT beam and $B_w$ is the WMAP beam. The calibration factor $\alpha$ appears in the cross and auto-spectra of ACT, and it can be estimated through relevant ratios of the auto- and cross-spectra. 
We define
\bea
\label{eqn:alphas}
\alpha_1(l) &=& C_l^{aw}/C_l^{aa}, \\ \nonumber
\alpha_2(l) &=& C_l^{ww}/C_l^{aw}, \\ \nonumber
\alpha_3(l) &=& \sqrt{(C_l^{ww}/C_l^{aa})}, \\ \nonumber
\eea
in which $C_l^{ww}$ is the  average $94$ GHz (W-band) power spectrum of the WMAP 7-year data  \citep{larson/etal:prep}\footnote{\url{http://lambda.gsfc.nasa.gov/product/map/dr3}} binned in the same way as the ACT auto- and cross-spectra.
The above three measures of the calibration (eqn. (\ref{eqn:alphas})) are not independent, but they can be used together to test the consistency of our results over a wide range of angular scales. 
Variation of $\alpha$ versus $l$ is a sign of scale dependent calibration factor. For the ACT \arone{} maps, the calibration factor is constant in the range of $400 < l < 1000$, and different measures are consistent with each other within the errors.  We restrict our analysis to this region. For $l\le 300$ the ACT data is dominated by the atmospheric noise. Beyond $l>1000$ WMAP maps are resolution limited.

  \begin{figure*}[htb]
\centering{ \resizebox{0.6\textwidth}{!}{
  \plotone{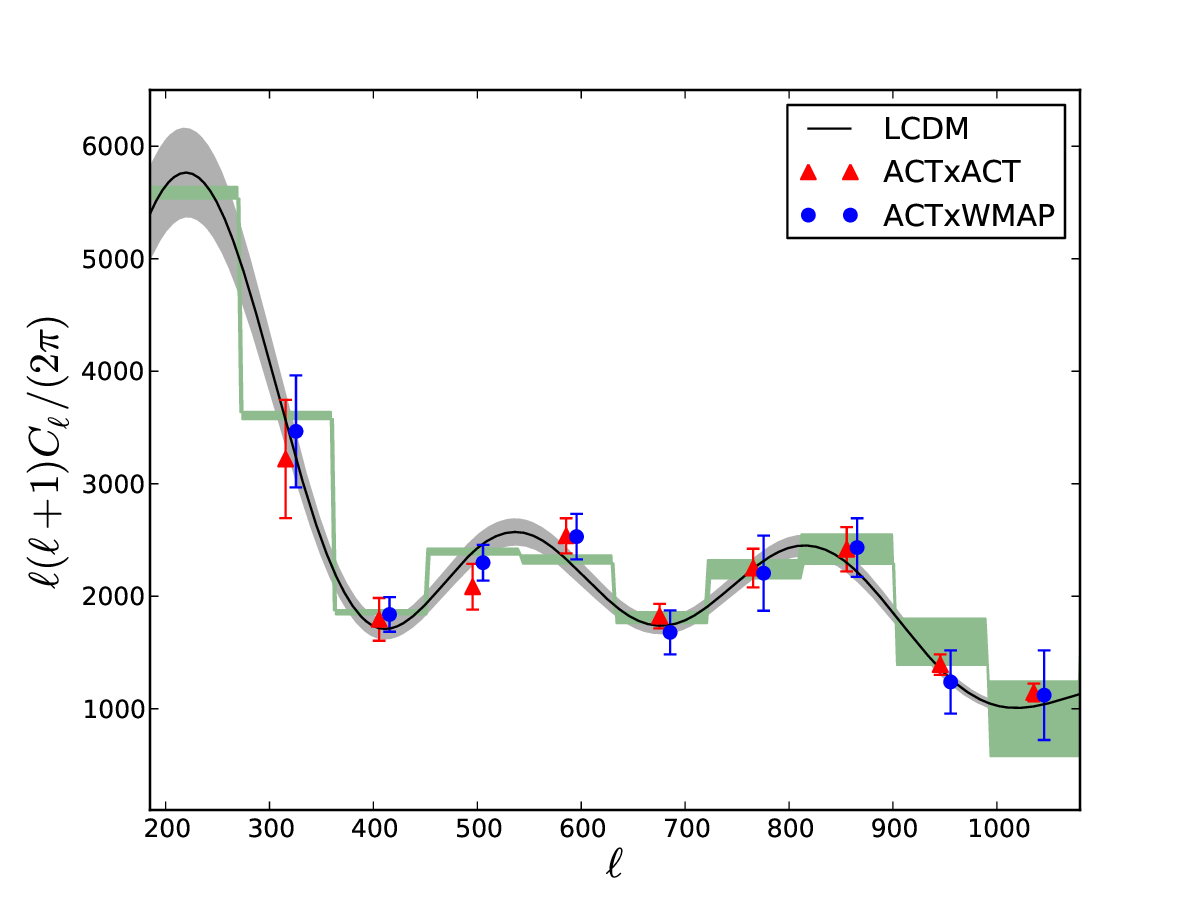}
  }}
  \caption{ ACT$\times$WMAP cross-power spectra. The ACT \arone{} - WMAP cross-spectrum is shown with blue dots and the single frequency spectrum of the ACT \arone{} maps is shown with red triangles. The average $94$ GHz (W-band) power spectrum of the WMAP data (light green squares) and the theory power spectrum based on WMAP7 best fit parameters (solid black line) are also plotted for reference. The gray band shows the cosmic variance.}
  \label{fig:calibrated}
\end{figure*}
The power spectrum bins are chosen such that the covariance between bins is negligible. Therefore the ratios of the spectra, $\alpha(l)$, provide independent measurements of the calibration factor at every $l$ bin. 
For the ACT \arone{} maps the $\alpha(l)$ are flat over the range of $400 < l < 1000$. Hence the overall calibration factor $\bar{\alpha}$ can be estimated from the average of the $\alpha(l)$ values in that $l$ range. Since the covariance between the $\alpha(l)$ values at different $l$ is negligible, the variance of the quantities that are used in averaging is a good measure of the error on the mean. Therefore we obtain:  
\bea
\bar{\alpha}_1 &=&  1.01 \pm 1.9\%, \\ \nonumber 
\bar{\alpha}_2 &=&  1.00 \pm 2.1\%,   \\ \nonumber  
\bar{\alpha}_3 &=&  1.00 \pm 1.4\%.  \\ \nonumber  
\eea

The uncertainties quoted are the error on the mean derived from the variance of the $\alpha(l)$ for every case. We use the Anderson-Darling statistic to test the normality of  the $\alpha(l)$ measures. The $A^2$ statistic that quantifies deviations from normality in this test becomes large when data points deviate from normality.
If $A^2 > 0.751$, the hypothesis of normality is rejected at the 95\% confidence level (for a 90\% confidence level it is 0.632). We compute $A^2$  for the three measures of the calibration factor defined above. The result is  $A^2(\alpha_1) = 0.27$, $A^2(\alpha_2) = 0.26$ and  $A^2(\alpha_3) = 0.43$. Therefore at the $>$10\% level, the data used in computing the average calibration factor do come from a normal distribution.

Among the above three measures of the calibration factor, $\bar{\alpha}_3$ has the smallest error. The reason is that it uses the sky averaged WMAP power spectrum as the estimate of the WMAP spectrum. Smaller errors in this quantity translate into the smaller error in the $\bar{\alpha}_3$. However all three measurements are in agreement with each other. For comparison, the Uranus calibration is $0.99 \pm 7.0\%$. We use 2\% as the calibration error for the \arone{} maps.

The calibrated power spectra are shown in Figure \ref{fig:calibrated}. The ACT \arone{} - WMAP cross-spectrum is shown with blue filled circles,  and single frequency spectrum of the ACT \arone{} maps is shown with red triangles. The average $94$ GHz (W-band) power spectrum of the WMAP data binned in the same way as other spectra are binned (light green boxes) and the theory power spectrum based on WMAP7 best fit parameters (solid black line) are also plotted for reference. ACT$\times$ACT power spectrum has large uncertainties on large scales due to the large scale noise in the ACT maps. WMAP power-spectrum has larger errors on smaller scales where WMAP detector noise dominates and maps are resolution limited. 

\subsection{Calibration: ACT \artwo{}}
We measure the calibration factor for the \artwo{} maps in a similar way. The result is a calibration factor with $\sim7\%$ fractional uncertainty in temperature. The larger uncertainty on the ACT \artwo{} calibration factor is due to the higher noise level in those maps than the \arone{} maps on large angular scales.
The calibration factors determined from cross-correlations with WMAP and from Uranus observations are consistent to $\sim7\%$. We use 7\% calibration error, based on Uranus, for the \artwo{} maps.

\begin{figure*}[htb]
\centering
    \resizebox{0.99\textwidth}{!}{
     \plotone{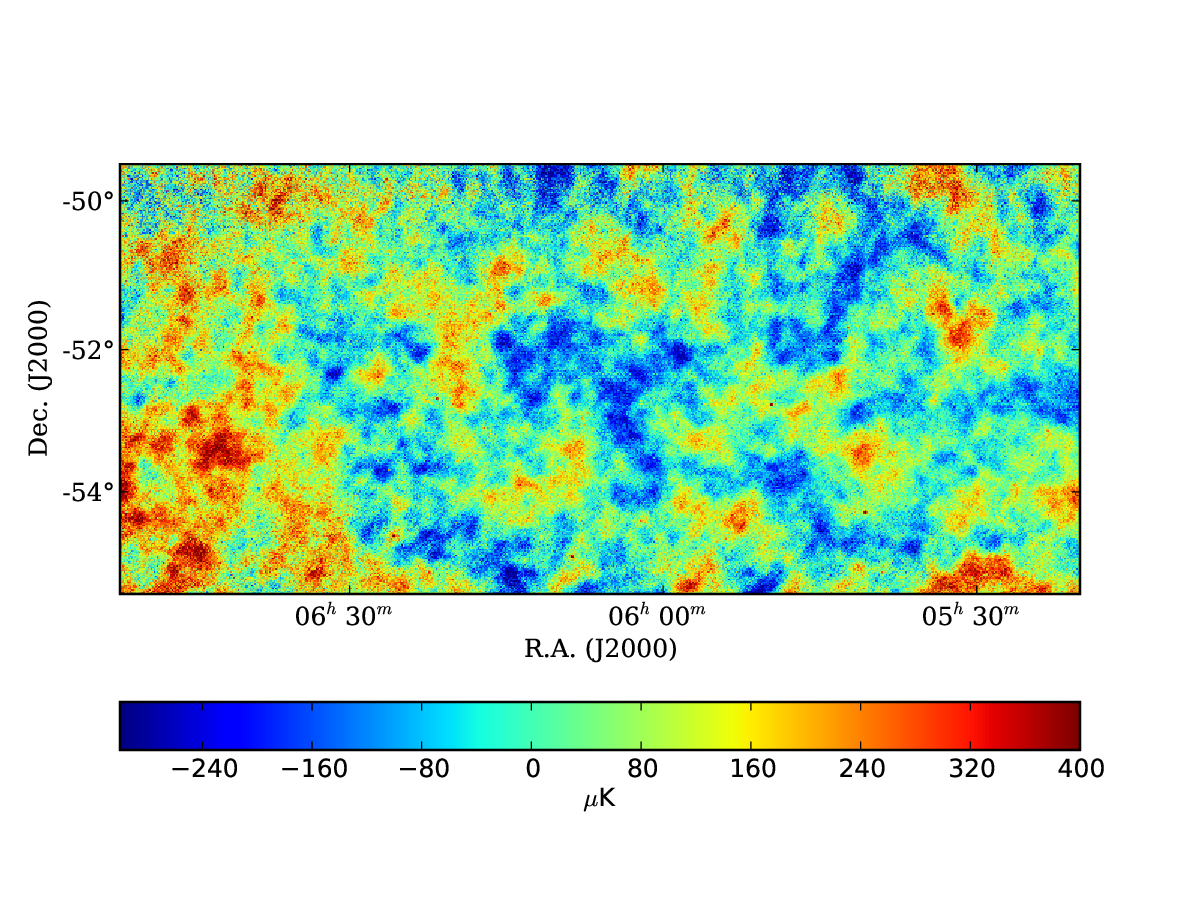}
}
  \caption{A 84 $\rm deg^2$ combined map obtained by combining WMAP and ACT \arone{} maps. This high resolution map encompasses all components of the microwave sky: the large scale structure is the CMB, hot (red) dots are point sources and some of the small cold (blue) spots are clusters of galaxies. }
  \label{fig:wMapPlusACT}
\end{figure*}
\subsection{Calibration systematics}
Several factors can systematically affect our results:
\begin{itemize}
\item{Stability vs. time and space:} In order to examine the stability of our calibration results in time and space, we compare the average calibration factors in the \npatch patches.  The average $\alpha_l$ in every patch is computed by averaging the calibration factor for different $l$ values in that patch. The scatter of the average calibration factors in \npatch patches gives us a measure of stability of our results versus the position in the sky and the time that the patch was observed. We verify that the patch-to-patch variation of the calibration factor is consistent with the 1$\sigma$ spread in the average calibration factor.
We have also checked the independence of the calibration factor on the area of the map. The calibration factor measured on the 228 $\rm{deg}^2$ data of \cite{fowler/etal:prep} is the same as that measured on the $\sim$300  $\rm{deg}^2$ map used in this work.
\item{Beam uncertainty and window function normalization:} As discussed in \cite{page/etal:2003}, limited knowledge of the beam profiles leads to uncertainties in the experimental window function. These uncertainties result in a distortion of the power spectrum, which in turn affects our estimation of the calibration factor. One of the uncertainties is the uncertainty in the normalization of the ACT window function, which appears as an overall change in the calibration factor. The beam transfer function we use is normalized to unity at $l = 700$ and thus has no uncertainty at that $l$.  The uncertainty in the beam transfer function is taken into account when estimating cosmological parameters \citep{dunkley/etal:prep} and we do not deal with it separately here.  
\item{Pointing reconstruction error:} The error due to the pointing reconstruction is discussed in \cite{jones/etal:2006} and causes a correlated distortion of the power spectrum at different bins. Absolute detector array pointings for the ACT data are established with $3.5''$ precision through an iterative process in which the absolute pointing is adjusted based on offsets of ACT-observed radio source locations with respect to source locations taken from the Australia Telescope 20 GHz (AT20G) survey  \citep{murphy/etal:prep}.
The bias induced by this error in the window function is not significant at the scales of our interest. 
\item{The band center uncertainty:}  Our calibration method is based on the CMB cross correlations in two different frequencies. The CMB spectrum in the frequency range of this work is flat and therefore uncertainty in the frequency band center of the experiment will not affect our results. 
\end{itemize}

\section{Combining Maps}
\label{sec:optimalMap}
The ACT and WMAP views of the CMB sky are complementary.  WMAP accurately measures large-scale CMB features and ACT provides a high resolution image of small scale features.  As we have shown in sections 3 and 4, these maps are consistent and now cross-calibrated so that they can be combined to make a CMB map that is signal-to-noise dominated over a wide range of scales. The WMAP data are signal dominated on large scales, $l < 548$ \citep{larson/etal:prep} and the detector noise dominates at smaller scales. The ACT data are dominated by the unconstrained modes on large scales that arise from a combination of instrument properties, scan strategy, and atmospheric contamination. The ACT data have a good signal-to-noise ratio on intermediate scales, $500<l<5000$, and become noise dominated beyond that.
 The basics of our method for combining WMAP and ACT is similar to the real-space method described in \cite{tegmark:1999}. Working in real space has the disadvantage of having to deal with large matrices for large maps. Instead, we work in Fourier space. 
We construct a linear combination of the two maps in Fourier space by inverse noise weighting them such that the less noisy map dominates at each scale (these maps are not high-pass filtered). 
The WMAP data has lower angular resolution than ACT and the pixel size of WMAP is much bigger than that of ACT as well. We first deconvolve the WMAP data to the ACT angular resolution in Fourier space. Then we use inverse variance weighting to combine the two maps. The construction is given by 
\be
M(\vec{l}) = \frac{\sigma_{ww}^2(\vec{l})a(\vec{l})}{\sigma_{aa}^2(\vec{l})+\sigma_{ww}^2(\vec{l})}+\frac{ \sigma_{aa}^2(\vec{l}) w(\vec{l})}{\sigma_{aa}^2(\vec{l})+\sigma_{ww}^2(\vec{l})},
\ee
where $\sigma_{aa}$ and $\sigma_{ww}$ are the ACT and WMAP noise spectra in two dimensional Fourier space. The above combination picks up WMAP contributions on large scales and ACT contributions on small scales and is signal dominated over a large range of $l$. The combined Fourier map, $M(\vec{l})$, is transformed back to real space. The ACT noise model, $\sigma_{aa}$, is computed in the same way as described above. WMAP noise is estimated by a white-noise, i.e. constant, before  beam deconvolution. For beam deconvolved WMAP maps, the noise model, $\sigma_{ww}$, is the WMAP noise multiplied by the two dimensional beam window function in Fourier space.
The resulting map is shown in Figure \ref{fig:wMapPlusACT}. This map encompasses all components of a high resolution CMB map (CMB on large scales, point sources and clusters on small scales) in one single map.

\section{Conclusion}

We have presented a method for calibrating ACT maps using cross-correlations with WMAP. The fractional uncertainty in the temperature calibration factor obtained in this paper is $\sim 2\%$ and the absolute calibration agrees well with that derived from Uranus observations. The calibration uncertainty is inversely proportional to the total area of the maps used for cross-correlations. Adding more area will decrease the uncertainty in the calibration factor. We have also presented fast methods for comparing and combining CMB data sets. Using these methods, we have combined ACT with WMAP data to construct a high resolution map that has a good signal-to-noise ratio over a wide range of angular scales.

\acknowledgements

This work was supported by the U.S. National Science Foundation
through awards AST-0408698 for the ACT project, and PHY-0355328,
AST-0707731 and PIRE-0507768. Funding was also provided by Princeton
University and the University of Pennsylvania.  The PIRE program made
possible exchanges between Chile, South Africa, Spain and the US that
enabled this research program.  Computations were performed on the GPC
supercomputer at the SciNet HPC Consortium.  SciNet is funded by: the
Canada Foundation for Innovation under the auspices of Compute Canada;
the Government of Ontario; Ontario Research Fund -- Research
Excellence; and the University of Toronto.

AH, VA, SD and TM were supported through NASA grant NNX08AH30G.  ADH
received additional support from a Natural Science and Engineering
Research Council of Canada (NSERC) PGS-D scholarship. AK and BP were
partially supported through NSF AST-0546035 and AST-0606975,
respectively, for work on ACT\@.   HQ and LI acknowledge partial support
from FONDAP Centro de Astrof\'isica.  RD was supported by CONICYT,
MECESUP, and Fundaci\'on Andes.  ES acknowledges support by NSF
Physics Frontier Center grant PHY-0114422 to the Kavli Institute of
Cosmological Physics. KM, M Hilton, and RW received financial support
from the South African National Research Foundation (NRF), the Meraka
Institute via funding for the South African Centre for High
Performance Computing (CHPC), and the South African Square Kilometer
Array (SKA) Project.  JD received support from an RCUK Fellowship.  RH
received funding from the Rhodes Trust.  We would like to thank Norm Jarosik for useful discussions and his contributions. SD acknowledges support from the Berkeley Center for Cosmological Physics.  YTL acknowledges support from the World Premier
International Research Center Initiative, MEXT, Japan. NS is supported by the U.S. Department of Energy contract to SLAC no. DE-AC3-76SF00515. 
We acknowledge the use of the Legacy Archive for Microwave Background Data Analysis (LAMBDA). Support for LAMBDA is provided by the NASA Office of Space Science.  The data will be made public through LAMBDA (\url{http://lambda.gsfc.nasa.gov/}) and the ACT website (\url{http://www.physics.princeton.edu/act/}). Some of the results in this paper have been derived using the HEALPix \citep{gorski/etal:2005} package.

\end{document}